\documentclass[%
 reprint,
 amsmath,amssymb,
 aps,
]{revtex4-2}

\usepackage{graphicx}
\usepackage{dcolumn}
\usepackage{bm}

\usepackage{epsfig} 
\usepackage{float}
\usepackage{xcolor}
\usepackage{epstopdf}
\usepackage[labelfont=bf,textfont=md]{caption}
\usepackage[hidelinks]{hyperref}
\usepackage{amssymb}
\usepackage{amsmath}

\newcommand{\cc}[1]{\textcolor{black}{#1}}
\newcommand{\dd}[1]{\textcolor{black}{#1}}
\newcommand{\ddthesis}[1]{\textcolor{black}{#1}}

\newcommand{\blue}[1]{\textcolor{black}{#1}}

\newcommand{\ddtwo}[1]{\textcolor{black}{#1}}

\newenvironment{ddenvironment}{\par\color{black}}{\par}

\begin{document}

\preprint{APS/123-QED}

\title{Inverse design of multishape metamaterials}

\author{David M.J. Dykstra}
\author{Corentin Coulais}%
 \email{coulais@uva.nl}
\affiliation{%
Institute of Physics, University of Amsterdam,  Science Park 904, 1098 XH, Amsterdam, the Netherlands\\
}%

\date{\today}

\begin{abstract}
Multishape metamaterials exhibit more than one target shape change, e.g. the same metamaterial can have either a positive or negative Poisson's ratio.  
%
{So far\ddtwo{,} multishape metamaterials have mostly been obtained by trial-and-error.} \ddtwo{The} inverse design of multiple target deformations in such multishape metamaterials remains a largely {open problem}. 
Here, we {demonstrate that it is possible to design metamaterials with multiple \cc{nonlinear} deformations of arbitrary complexity. To this end, we}
introduce a novel sequential nonlinear method to design multiple target modes. We start by iteratively adding local constraints that match a first specific target mode; we then continue from the obtained geometry by iteratively adding local constraints that match a second target mode; and so on. We apply this sequential method to design up to 3 modes with complex shapes and we show that this method yields at least \ddtwo{an} 85\% success rate. \dd{Yet we find that these metamaterials invariably host additional spurious modes, whose number grows with the number of target modes and their complexity, as well as the system size.} 
Our {results} highlight an inherent trade-off {between design freedom and design constraints and pave the way towards } 
multi-functional materials and devices.
\end{abstract}

\maketitle


\emph{Introduction.} ---
Designing for multiple objectives is a notoriously difficult computational task. The particular challenge of designing complex on-demand shape changing structures and metamaterials is especially acute because shape-changing typically is a nonlinear problem~\cite{Bertoldi_NatRevMat}. 
Over the past few years, many design methods have been introduced to create shape-shifting kirigami~\cite{Sussman_PNAS2015,Celli_SoftMatter2016,choi2019programming,choi2021compact,dudte2022additive,Jin_AdvMat2020}, origami~\cite{dudte2016programming,dudte2021additiveorigami} or cellular metamaterials~\cite{coulais2016combinatorial,Siefert_NatMat2019,tricard_ACM2020,Czajkowski_NatComm2022}. 
%
Among these inverse design methods, two main analysis methods can be distinguished: elastic analyses~\cite{ronellenfitsch2019inverse,oliveri2020inverse,kumar2020inverse} and mechanism-based analysis with zero-energy modes~\cite{coulais2016combinatorial,choi2019programming,kim2019conformational,dudte2022additive}. 
On one hand, inverse design of mechanical metamaterials using elastic methods can be done in a variety of ways including machine learning~\cite{kumar2020inverse} and topology optimization~\cite{xu2022inverse}, which can even be done for multiple modes simultaneously~\cite{oliveri2020inverse,ronellenfitsch2019inverse}. \dd{However,} {such methods are typically limited to small deformations and are more difficult to expand to shape-changing structures.}
On the other hand, 
mechanism-based inverse analysis methods often assume unit cells with a single degree of freedom, such as four-bar linkages in 2D~\cite{choi2019programming,choi2021compact,ou2018kinetix} and 3D~\cite{coulais2016combinatorial,ou2018kinetix} or origami patterns with quadrilateral faces~\cite{dudte2016programming,Dieleman_NatPhys2020,dudte2021additiveorigami}. {Crucially, mechanism-based metamaterials naturally lead to large deformations and shape-morphing. Yet such} metamaterials have \cc{mostly} been designed {with a single shape-change only. \cc{The rare examples of} inverse design of multiple {shapes} \cc{have achieved 2 shapes only with some limitations on the complexity of the 
shape-changes~\cite{kim2019conformational,Dieleman_NatPhys2020}.}

\begin{figure}[!b]
\includegraphics[width=1.\columnwidth]{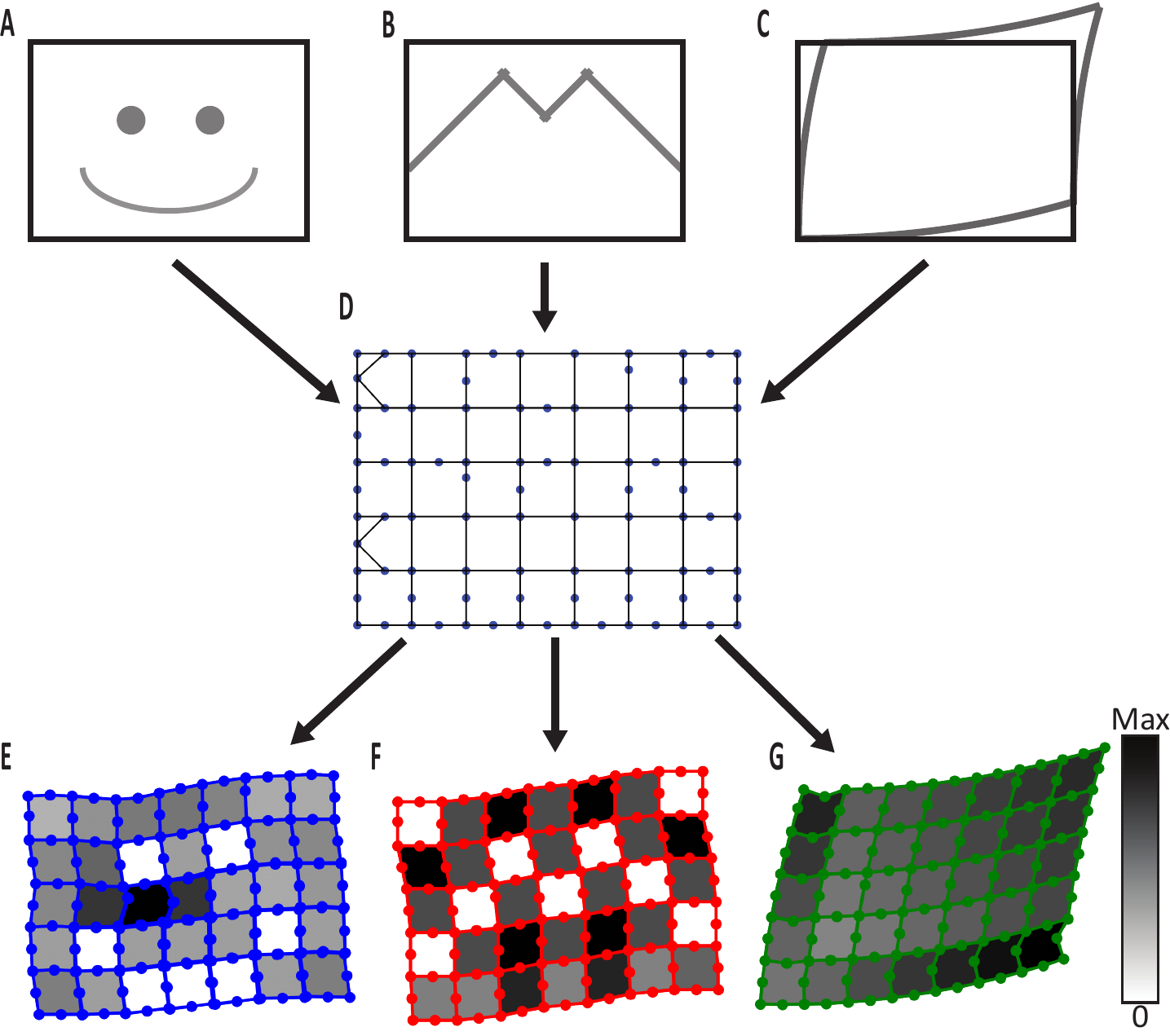}
\caption{\textbf{Inverse Multimodal Design.} (A-C) A base rectangular mechanical metamaterial (black) has three target deformation modes (gray): (A) a mode where everything except a smiley deforms, (B) a mode where everything except an M deforms, (C) a curving mode. (D) The $7 \times 5$ lattice with hinges in blue and bars in black accommodates all modes of (A-C) in (E-G) respectively. Colors in gray indicate the sum of the absolute values of the angular deformations per unit cell, with no deformation in white and maximum deformation per mode in black, according to the color bar.}
\label{fig1}
\end{figure}

Here, we introduce a sequential design method for multiple modes in mechanism-based metamaterials~\cite{milton2013adaptable,lubbers2019excess,bossart2021oligomodal,bossart2022extreme,van2022machine}. We construct an iterative algorithm at the level of each unit cell to design one target mode, which we in turn apply in sequence to multiple target modes. We find that this algorithm almost always succeeds in creating metamaterials with multiple on-demand modes, but that twice as many spurious modes emerge during the design process. Our sequential method establishes a first foray into the design of multiple \cc{complex} deformation modes and hints at an interplay between target number of modes and total number of modes. \cc{Our method is complementary to combinatorial approaches~\cite{coulais2016combinatorial,Meeussen_NatPhys2020,Dieleman_NatPhys2020,bossart2021oligomodal,van2022machine,bossart2022extreme}. These methods are often constrained in their design freedom and constrained mechanically, whereas our method starts from a highly unconstrained geometry that provides greater design freedom but is less constrained mechanically. }More broadly, our study opens up a promising avenue for the design of multifunctional materials.

\emph{Inverse design of multiple modes.} ---
An example of such inverse design can be seen in Fig. \ref{fig1}. First, we select three desired modes of deformation in Fig. \ref{fig1}ABC respectively. Coulais et al.~\cite{coulais2016combinatorial} showed that combinatorial designs could be used to generate any desired texture in a metamaterial, such as a smiley. In Fig. \ref{fig1}A, we choose an opposite target mode instead: a mode where all unit cells deform except for an undeformed smiley face in gray. Similarly, for a second target mode in Fig. \ref{fig1}B, we choose to deform all unit cells instead of a central \textit{M} for \textit{Metamaterials}. Finally, for a third and final mode in Fig. \ref{fig1}C, we choose a global \ddtwo{curved} shear mode. Using our inverse design algorithm, we can translate these three target modes into the bar-node mechanism of Fig. \ref{fig1}D. We can then deform this mechanism to display the modes of Fig. \ref{fig1}EFG, where a \ddtwo{white} color highlights an undeformed cell. As expected, Fig. \ref{fig1}EFG correspond very well with the target modes of Fig. \ref{fig1}ABC. In fact, Fig. \ref{fig1}C, which shows large nonlinear deformation, demonstrates that the algorithm also works succesfully for large deformations. On the other hand, Fig. \ref{fig1}AB show little deformation because small deflections are used. However, these deformations can be visualized very well using a vertex representation, which will be explained in the next section.

\emph{Unit cell definition.} ---
To explain how the algorithm works, we start from a square base cell. In Fig. \ref{fig2}A, a square base cell with four nodes is shown. This base cell has a single mode of deformation, shown in Fig. \ref{fig2}D, which would host a mode of counter-rotating squares in a periodic tiling~\cite{grima_auxetic}. This metamaterial design features a negative Poisson's ratio and has been widely explored~\cite{grima_auxetic,Bertoldi_NatRevMat}. However, within the same square domain, we can define a wide variety of other mechanisms, such as the distorted hexagon and regular octagon in Fig. \ref{fig2}BC respectively. A possible form of deformation for each polygon can be seen in Fig. \ref{fig2}DEF respectively. For small deformations, it can be difficult to visualize the actual deformations. However, the deformations can also be represented using a vertex representation as has been done in Fig. \ref{fig2}GHI~\cite{bossart2021oligomodal,van2022machine}. Here, vertices originating from the centre of the base cell are seen representing the the nodes of the polygon. Fig. \ref{fig2}JKL then add arrows on top, where the area of the arrows corresponds to the corresponding angular hinge deformation~\cite{bossart2021oligomodal,van2022machine}. Just as the sum of the angular deflections in closed polygons must add up to zero, the sum of the arrows, weighed by their size is zero, in what is also known as the ice rule in the context of vertex models~\cite{pauling1935structure}. Two neighboring vertices must have an equal angular deflection, viz. share the same arrow.
This vertex representation is a convenient way to represent
angular deformation and determine compatibility constraints. {Crucially, we enforce such compatibility constraints at the nonlinear level up to quadratic order, see} \ddtwo{Appendix \ref{subsec:solbasecell}} {for a discussion of the linear order and } \ddtwo{Appendix \ref{subsec:solpoly}} { for a detailed discussion of the quadratic order.}


\begin{figure}[t!]
\includegraphics[width=1.\columnwidth]{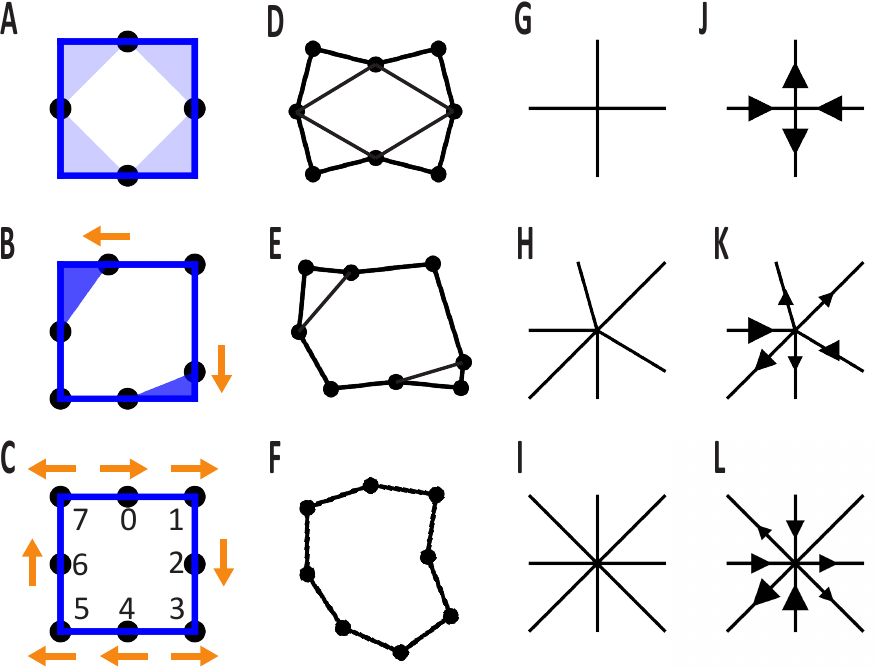}
\caption{\textbf{Base cell geometry.} A square lattice can host a variety of polygons including (A) a square, (B) a distorted hexagon and (C) an octagon. Black dots indicate hinges. Numbers in (C) correspond to node numbers, while the orange arrows indicate nominal directions of distortion. Possible deformation modes of (A-C) include (D-F) respectively. The undeformed unit cells (A-C) can be represented using the representation of (G-I), where arrows can be added in (J-L) to represent the deformations of (D-F) respectively~\cite{bossart2021oligomodal}.}
\label{fig2}
\end{figure}

\begin{figure*}[t!]
\includegraphics[width=1.\textwidth]{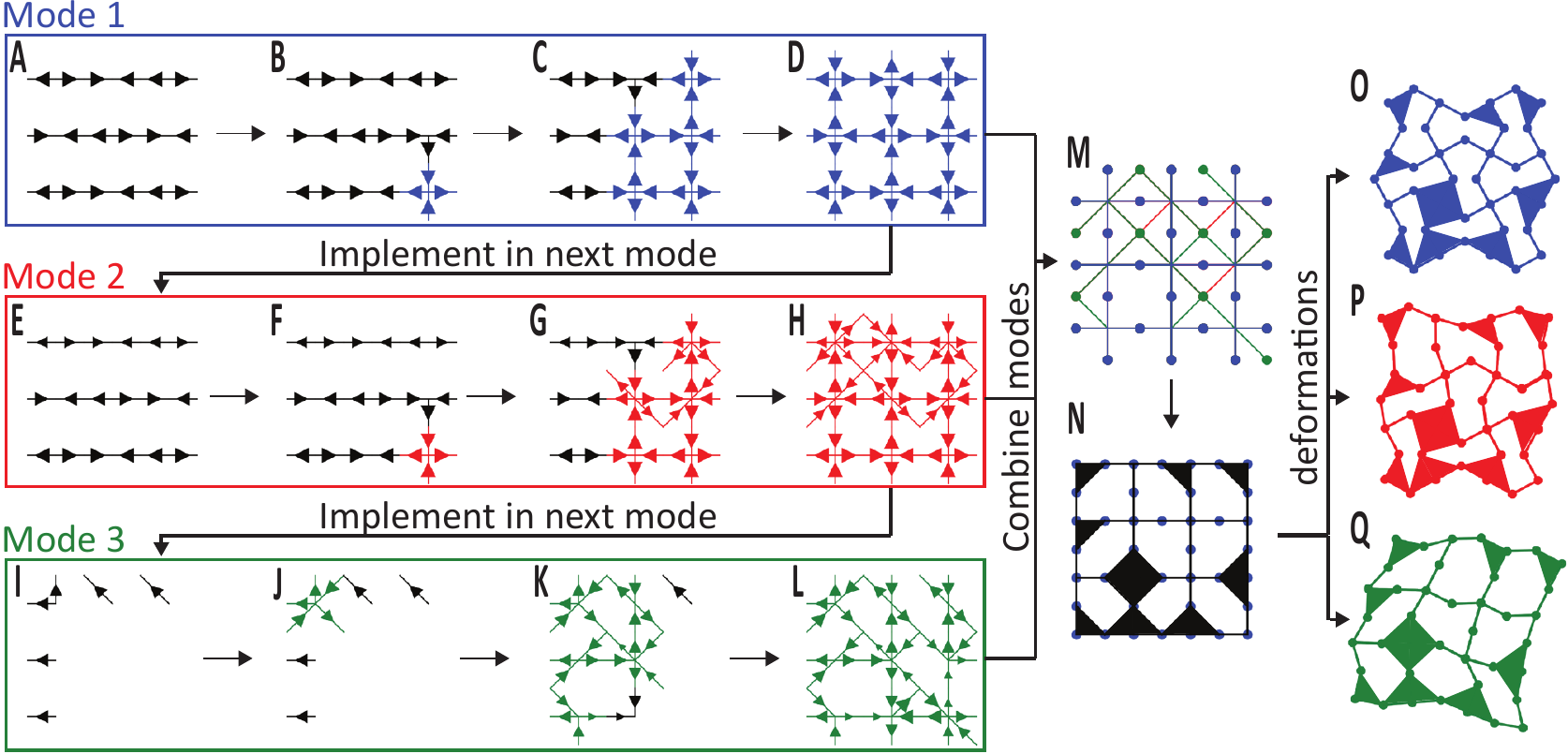}
\caption{\textbf{Sequential Design Algorithm.} A $3 \times 3$ metamaterial has three target modes, with local vertex deformations defined in (A,E,I). From target mode one (A), fitting base cell deformations are implemented one by one (blue (B,C)) until the mode is fully solved (D). All vertex node locations of  (D) are implemented in the start vertices of target mode two (E), after which fitting unit cells are implemented again in (F-H). This process is repeated for target mode three (I-L). The vertex locations of (D,H,L) are combined in (M), which corresponds to the mechanism in (N). The mechanism in (N) can host all physical deformations of (O,P,Q) which correspond to the vertex deformations of (D,H,L) respectively.}
\label{fig3}
\end{figure*}

\emph{Sequential design algorithm. ---}
We can then use this method to solve unit cells to inversely solve desired metamaterial patterns, as illustrated in Fig. \ref{fig3}. In this example, we have a $3 \times 3$ mechanical metamaterial with 3 vertex target patterns in Fig. \ref{fig3}AEI respectively. We solve these modes sequentially as follows:

\begin{enumerate}
    \item We start from desired mode 1 deformations in Fig. \ref{fig3}A and
    fit compatible unit cells one at a time (blue)
    (\ref{fig3}B-D). 
    When selecting those unit cells, we follow three important criteria: 
    (a) we always solve the unit cell with the largest number of defined vertices~\footnote{If several unit cells have the two vertices defined In (Fig. \ref{fig3}A), we pick a cell randomly, In Fig. \ref{fig3}B, the cell above the previously solved unit cell then has more vertices defined and will be selected next to be solved.}. 
    (b) We select solutions that have the lowest possible number of degrees of freedom. 
    (c) We select solutions that have the smallest variation possible of the size of the deformation.
    \item Once we have selected all the unit cells, the mode is solved in Fig. \ref{fig3}D and we continue with mode 2.
    The vertex locations of the solution of mode 1 of Fig. \ref{fig3}D are combined with the input deformation of Fig. \ref{fig3}E. Mode 2 (red) is solved sequentially in Fig. \ref{fig3}E-H in the same way as mode 1 was.
    \item The approach above is repeated in Fig. \ref{fig3}I-L to obtain mode 3 (green). In principle, the approach could be repeated to allow for more modes, as long as sufficient unspecified degrees of freedom remain.
    \item All degrees of freedom of all modes are combined in a single vertex representation in Fig. \ref{fig3}M.
    The vertex representation of Fig. \ref{fig3}M can be translated to the mechanism design of Fig. \ref{fig3}N. Finally, the vertex representation of the three modes in Fig. \ref{fig3}DHL, can be translated to the real deformations of Fig. \ref{fig3}OPQ respectively.
\end{enumerate}

{We have used this algorithm for creating the geometry in Fig.~\ref{fig1}, see also Fig. \ref{fig_solve7x5} for a graphical representation of the algorithm and Appendix} \ddtwo{\ref{subsec:how_algo_7x5}} {for details.}
{Importantly}, all of the modes of Fig. \ref{fig1} and Fig. \ref{fig3} show large deformations. This highlights a distinct advantage of our nonlinear method, which could not have been achieved with a linear method. 
In \ddtwo{Appendix \ref{subsec:Energy}}, we show that our quadratic nonlinear solution can be up to eight orders of magnitude more accurate than a linear solution in solving a unit cell with small displacements. 

\emph{Spurious modes. ---}
While {our method is successful at} creating metamaterials that can host three {on-demand} modes in Fig. \ref{fig1} and Fig. \ref{fig3}, we do not yet know whether any spurious modes are created in the process. This is important because a metamaterial with many modes will be more difficult to actuate than a metamaterial with less modes: it will require a more specific loading that only actuates the modes of interest and its response will be less robust. For this reason, we calculate the linear number of modes of our solved solutions. To do so, we first construct the compatibility matrix of each design, such as those in Fig. \ref{fig1}D and Fig. \ref{fig3}N. We then calculate the dimension of the kernel of the compatibility matrix~\footnote{To this end, we use QR factorization, as implemented in the \textit{linalg} package from \textit{scipy} in Python. The number of modes is then equal to this dimension including three rigid body modes: two translational and one rotational~\cite{van2022machine}.}. We then find that the design of Fig. \ref{fig1} features significantly more than three modes, namely 47. 
However, if we solve only for the smiley of Fig. \ref{fig1}E, we find 9 modes. When we include the \textit{M} as a second mode, we in turn find that the lattice features 17 modes.Therefore, although one is able to create a metamaterial with up to three on-demand modes of arbitrary complexity, one systematically ends up with a large number of spurious modes. Interestingly, the more on-demand modes we require, the larger the number of spurious modes. \cc{This is presumably the flip side of our method. Since the procedure starts from an under-constrained lattice, it is very successful for the design of arbitrarily complex modes, but as a result of this design freedom the procedure is not able to constrain unwanted modes.}

\begin{figure}[t!]
\includegraphics[width=1.\columnwidth]{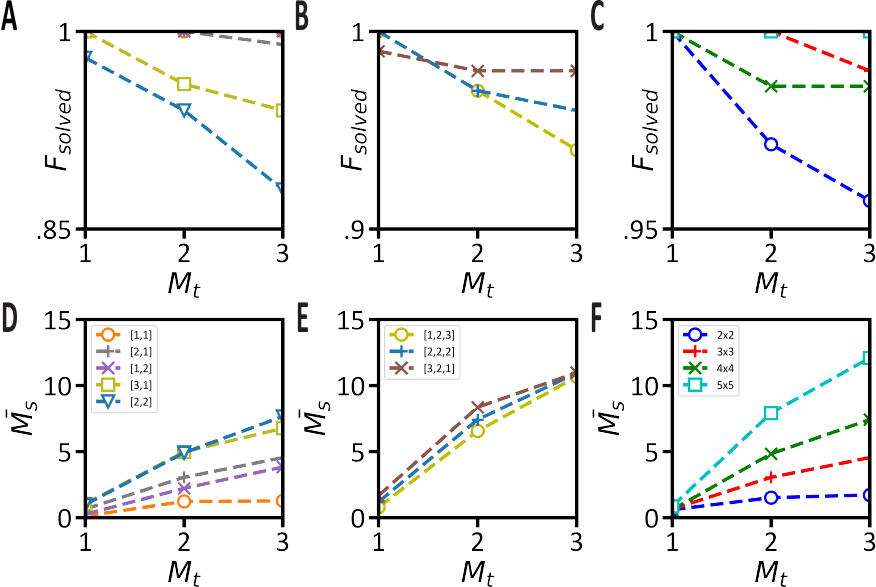}
\caption{\textbf{Statistical analysis of inverse design.} As function of the target number of modes, $M_t$, \ddtwo{(ABC)} the fraction of cases in which a solution was found, $F_{solved}$, and \ddtwo{(DEF)} the mean number of spurious modes, $\bar{M_s}$ 
\ddtwo{(AD)} shows the effects of input complexity on a $3 \times 3$ lattice, with $[i_c, i_v]$ = [number of cells on which inputs are defined, number of input vertices defined per input cell. \ddtwo{(BE)} shows the effects of order of complexity on a $4 \times 4$ lattice, with $[i_{c,1}, i_{c,2}, i_{c,3}]$ the number of cells on which inputs are defined for mode 1, 2 and 3 respectively, each which $i_v=1$. \ddtwo{(CF)} shows the effects of system size, with $N \times N$ the system size and $[i_c, i_v] = [2,1]$.}
\label{fig4}
\end{figure}

\emph{Statistical analysis. ---}
To further investigate {this emerging interplay between target and spurious modes}
, we systematically run many inverse design analyses with a variety of random input deformations: 100 analyses per analysis variation. We track the performance of our algorithm in Fig. \ref{fig4}. 
In  Fig. \ref{fig4}{ABC} we track how often our algorithm succeeds in finding a solution as function of the variations. We find for all cases that our method is able to find a solution in more than 85 \% of the cases, even when when we require three target modes of complex shape. This shows that our method is consistent and reliable.

In  Fig. \ref{fig4}{DEF} we analyze how the mean number of spurious modes, $\bar{M_s}$, develops as function of the variations~\footnote{The number of spurious modes, $M_s=M_P-M_t$, with $M_P$ the produced number of modes and $M_t$ the target number of modes. $M_s$ cannot be negative.}. In all cases, we find that the average number of spurious modes {grows} sublinearly with the target number of modes. Crucially, the higher the complexity of the target modes, the larger the number of spurious mode $M_s$ (Fig. \ref{fig4}{D}). Indeed, more complex modes require more design freedom and this additional freedom comes at the cost of additional spurious modes, as well a reduced chance of solving (Fig. \ref{fig4}A). Does it then matter whether we solve simple or complex modes first? We answer this question with the help of Fig. \ref{fig4}{BE}, where we vary the complexity between the three target modes either from simple to difficult (green $\circ$), with a constant difficulty (blue $+$) or with a decreasing difficulty (brown $\times$).  In Fig. \ref{fig4}{E}, we see that at $M_t=3$, the order of difficulty is irrelevant for the number of spurious modes. However, we do see in  Fig. \ref{fig4}{B}, that the chance of finding a suitable solution for all three modes increases when solving from  difficult to easy. {Therefore, finding a complex solution is more likely with more design freedom.} \ddtwo{This approach was also used in Fig. \ref{fig1}, where we solved the modes in the order of decreasing difficulty.}

{Additional design freedom is also provided by larger system sizes. Indeed, we see in Fig. \ref{fig4}C, that while the design of $2\times 2$ systems has always a $95\%$ success rate, $5\times 5$ systems reach a $100\%$ success rate.} \ddtwo{An increase in system size however, typically comes hand in hand with more spurious modes, Fig. \ref{fig4}}{F}.

\begin{figure}[t!]
\includegraphics[width=1.\columnwidth]{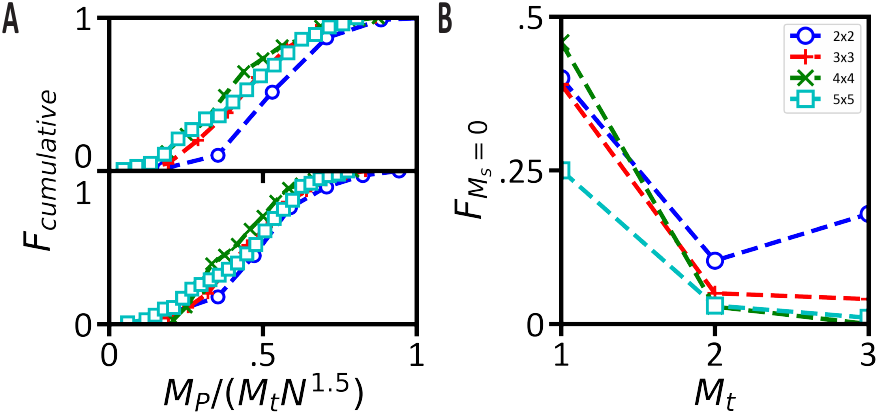}
\caption{\textbf{Modal distribution.} \ddtwo{(A)} The cumulative distribution, $F_\textrm{cumulative}$, of the number of modes, $M_P = M_t+M_s$, normalized by $M_t N^{1.5}$. Top: $M_t=2$, bottom: $M_t=3$. 
\ddtwo{(B)} Fraction of cases without spurious modes, $F_{M_s=0}$, as number of the target number of modes, $M_t$ for various system sizes. }
\label{fig5}
\end{figure}

{This competition between design freedom and constraints can also be clearly seen by considering the cumulative distribution of the total number of modes $M_P=M_t+M_s$ rescaled by $M_t\times N^{1.5}$, Fig.~\ref{fig5}A. We immediately see that such rescaling leads to a reasonable collapse of all the cumulative distributions. This collapse means that designs with more design freedom---they either require more target modes or are performed in larger systems---not only have a larger total number of modes, but they also have a broader} \ddtwo{mode number} distribution.

{Finally, that the distribution broadens also suggests that it is possible to achieve no spurious modes. To investigate this issue, we plot}
the fraction of cases, where there are no {spurious} modes in Fig. \ref{fig5}{B}. For a single mode, our inverse design method generates a single output in more than 25\% of the cases, regardless of system size. This shows that our method can be used for inversely generating metamaterials with a single output mode. For two or three target output modes, this fraction drops rapidly. Therefore, while it is possible to design large metamaterials with multiple target modes, it is highly unlikely, at least with our method in its current form. \cc{It remains an open question whether one could keep the design freedom without constraining the spurious modes.}

\emph{Discussion. ---}
Inversely designing metamaterials with multiple target deformations can be very complex and many approaches can be chosen. While various optimization strategies are effective for linear systems, nonlinear cases are much more difficult. In this paper, we have introduced a sequential nonlinear method to do so by locally designing octagonal base cells on a square lattice to match the surrounding deformations with minimal flexibility. We have used this method effectively to create mechanical metamaterials which can simultaneously exhibit multiple complex target deformations. Moreover, while we have restricted ourselves to octagonal cells on a square lattice in this manuscript, our method is general by nature and can be applied to any two-dimensional lattice with polygons of arbitrary complexity {(see Appendix \ref{subsec:hexcell} for an example of a hexagonal unit cell)}. Furthermore, because this method uses localized design, earlier made designs can easily be expanded or merged together using more local design. 

\ddthesis{The method we have introduced belongs to a class of direct methods for inverse design as opposed to most inverse design approaches, which use an iterative global optimization process where local design keeps being adjusted to best fit the global design parameter~\cite{choi2019programming,oliveri2020inverse,ronellenfitsch2019inverse}. Using local unit cell design or selection instead of simultaneous design has distinct advantages in readjusting to different inputs and allows a linear scaling of the computational time with system size, instead of slower than linear for most global optimizers, such as finite element codes with topology optimization~\cite{bendsoe2003topology}. Furthermore, while many direct inverse design methods use combinatorial approaches~\cite{coulais2016combinatorial}, the present method offers an advantage of flexibility compared to combinatorial approaches, which {are} inherently {limited to} discrete solutions.}

However, while our method has shown itself effective to accommodate multiple target modes simultaneously, it has also shown that it is difficult to do so without introducing {spurious}
 modes. A significant challenge remains; either on how to restrict these additional modes afterwards or to prevent these {spurious} modes during the inverse design altogether. \ddthesis{Nevertheless, the current method has demonstrated that it can be used to obtain large deformations. While additional linear modes arise in the design, this does not imply that it remains easy to actuate these additional modes at large deformations. It is worth exploring the competition in strain energy between the desired and unwanted additional modes. Consequently, it is worth exploring how these modes could be excited in real metamaterials.}
 
 Finally, while the method presented is very general for any two-dimensional polygon, the method can still be expanded towards (quasi-)periodic and three-dimensional metamaterials. We anticipate applications in soft robotics, phononic and acoustic wave manipulation and multi-functional materials and devices.

\emph{Data and Code Availability.} ---
The data and codes that support the figures within this paper are publicly available on a Zenodo repository~\cite{Zenodo2023oligomodal}.

\emph{Acknowledgements.} ---
We thank Sebastiaan Kruize for preliminary work, and Ryan van Mastrigt and Martin van Hecke for discussions. We acknowledge funding from the European Research Council under grant agreement 852587 and the Netherlands Organization for Scientific Research under grant agreement NWO TTW 17883.

\bibliography{references}

\clearpage

\begin{appendix}


\section{Solving a base cell}
\label{subsec:solbasecell}

The shape of any two-dimensional polygon in turn can be described by three equations~\cite{peacock1845treatise}. To a \dd{linear} degree, any variation of the base cell, described by $a_0 - a_7$ and $d_0 - d_7$ can be described by the following three equations:
\begin{equation}
0= - a_{0} - \frac{3 a_{1}}{2} - a_{2}- \frac{a_{3}}{2}  + \frac{a_{5}}{2}	- \frac{a_{7}}{2} + d_{1}  + d_{3},
\label{eq:base1}
\end{equation}
\begin{equation}
0= + {a_{0}} + {a_{1}}- {a_{3}} - {a_{4}} -  {a_{5}}  + a_{7},
\label{eq:base2}
\end{equation}
\begin{equation}
0 = - {a_{0}} - {\frac{a_{1}}{2}}+ {\frac{a_{3}}{2}}  - {\frac{a_{5}}{2}} - {a_{6}} - {\frac{3 a_{7}}{2}} + {d_{5}} + {d_{7}}.
\label{eq:base3}
\end{equation}

The derivation of Eq. \eqref{eq:base1} to Eq. \eqref{eq:base1} is given in Section \ref{subsec:App_Deriv_oct}.  \dd{It can be observed that  Eq. \eqref{eq:base1} to Eq. \eqref{eq:base1} do not depend on $d_{0,2,4,6}$. This implies that $d_{0,2,4,6}$ do not affect the design in a linear analysis. This is because $d_{0,2,4,6}$ are in-line with the bars attached to nodes 0,2,4 and 6 respectively. In order to identify the effects of $d_{0,2,4,6}$, a minimum of a quadratic nonlinearity is required. The quadratically nonlinear versions of Eq. \eqref{eq:base1} to Eq. \eqref{eq:base1} are provided in Eq. \eqref{eq:base1quad} to Eq. \eqref{eq:base3quad} respectively. Moreover, implementing a quadratic nonlinearity yields more accurate results, as described in \ddtwo{the Main Text.}
Therefore, we opt to use quadratic nonlinearity throughout this article.}

As the geometry is defined by three equations, this implies that:
\begin{equation}
j=c-3,
\end{equation}

This also implies that if a set of deformations $a_i$ and distortions $d_i$ are known in advance, a fitting polygon can be plugged in with three additional deformations and distortions. For example, for Fig. \ref{fig2}B, if $a_{1,2,4,5}$ and $d_{1,2,4,5}$ are known in advance, it is possible to solve for $a_{0,6}$ and $d_0$, by solving Eq. \eqref{eq:base1} to Eq. \eqref{eq:base3}. We do this numerically using the \textit{findroot} algorithm implemented in the \textit{mpmath} package in Python. Since only three quadratic equations are solved, more efficient (quasi-)analytical solutions can also be adopted~\cite{sturmfels2002solving}.

The base cell can be generalized using the octagon of Fig. \ref{fig2}C. 
The variations of the unit cell are chosen to have a maximum of eight hinges or nodes, numbered clockwise from 0-7. The undistorted distance between each node is considered 1, giving an overall square size of $2 \times 2$. 
Each of these nodes can be shifted by a distance $d_i$ with respect to the undistorted cell in the direction of the orange arrows in Fig. \ref{fig2}C. For the hexagon of Fig. \ref{fig2}BH, this implies that $d_0 = -0.28$ and $d_2 = 0.6$. Furthermore, each angle can deform with an angle $a_i$. For the square of Fig. \ref{fig2}DJ, this implies that $a_0=a_4 = 0.5$ [rad] and $a_2=a_6 =-0.5$ rad.

\smallskip
\section{How the algorithm works on $7 \times 5$ samples}
\label{subsec:how_algo_7x5}

\begin{figure*}[ht!]
\includegraphics[width=1.\textwidth]{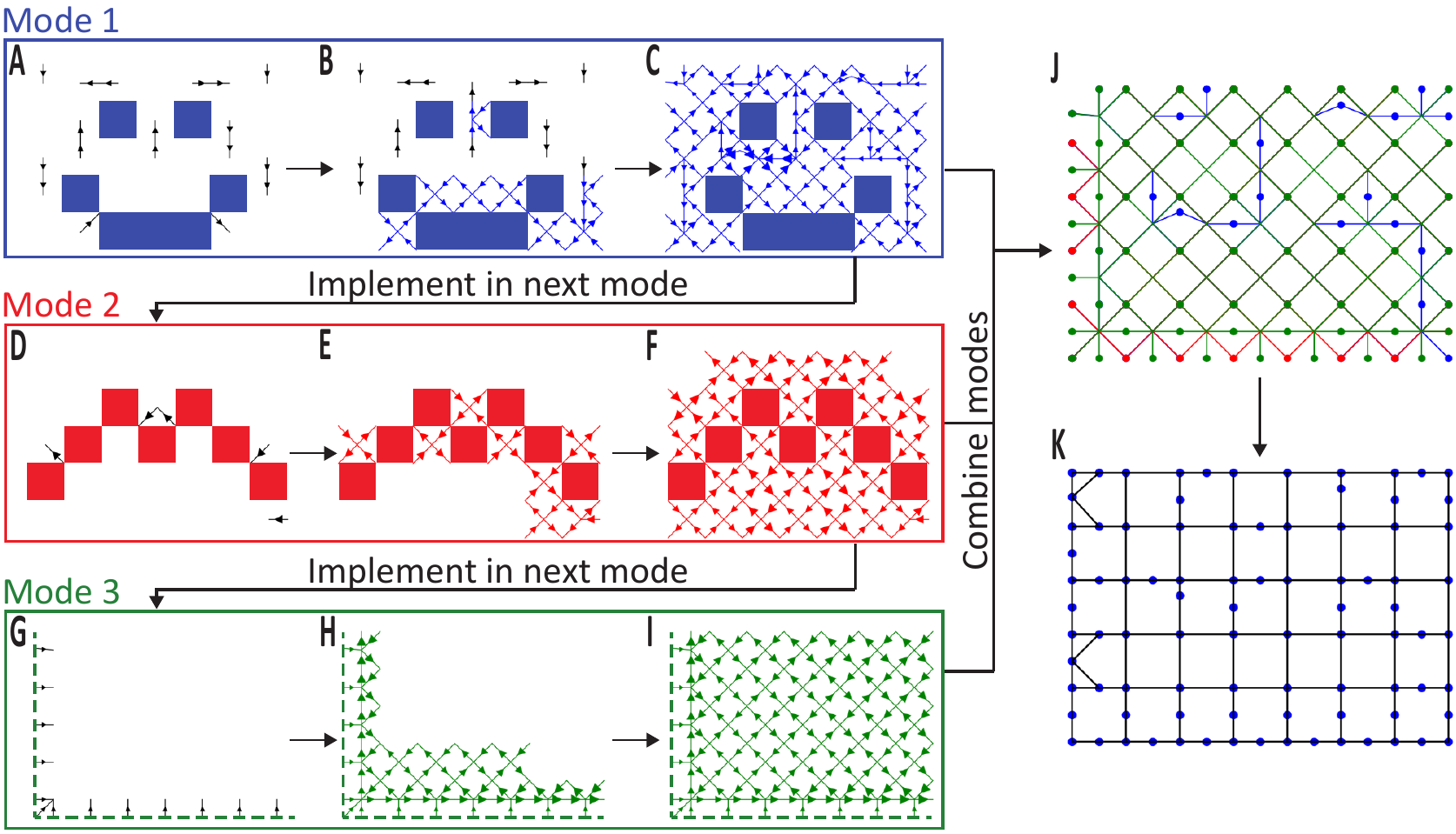}
\caption{\textbf{Inverse Design Algorithm.} A $7 \times 5$ metamaterial has three target modes, defined Fig. \ref{fig1}, with local vertex deformations defined in (A,D,G). Unit cells defined to be undeformed in (A,D) are highlighted with solid squares. Vertices along the perimeter defined to be undeformed in (G) are highlighted with a dashed line, unless the vertex representation indicates otherwise. From target mode one (A), fitting base cell deformations are implemented one by one until the mode is fully solved (blue (B,C)) . All vertex node locations of  (C) are implemented in the start vertices of target mode two (D), after which fitting unit cells are implemented again in (E,F). This process is repeated for target mode three (G-I). The vertex locations of (C,F,L) are combined in (J), which corresponds to the mechanism in (K). The mechanism in (N) can host all physical deformations of Fig. \ref{fig1}E-G which correspond to the vertex deformations of (C,F,I) respectively.}
\label{fig_solve7x5}
\end{figure*}

\ddtwo{Fig. \ref{fig3} and the accompanying text} showed how the algorithm works to solve for three modes in a relatively small sample of $3 \times 3$ unit cells. The following section shows the same process for the larger $7 \times 5$ metamaterial introduced in Fig. \ref{fig1}. This process can be seen in Fig. \ref{fig_solve7x5}. Furthermore, this section demonstrates how a target mode can be translated to input vertices.

First, the target modes of Fig. \ref{fig1}A-C are translated to the inputs defined in Fig. \ref{fig_solve7x5}ADG respectively. As we target an undeformed smiley in mode 1 and an undeformed \textit{M} in mode 2, we define those areas to be undeformed in Fig. \ref{fig_solve7x5}AD respectively. We then define a number of deformations around these undeformed areas to trigger deformations throughout the rest of the metamaterial. 

For mode 3, defined in Fig. \ref{fig1}C, we target a global deformation mode instead. As such, we define the entire bottom and left perimeter in Fig. \ref{fig_solve7x5}G instead. We do so by stating that the central vertex along this perimeter moves inwards for every unit cell along the perimeter, while we keep the other vertices on this perimeter undeformed. 

Using these inputs for the three defined modes, we can solve for the metamaterial design of Fig. \ref{fig1} as described in \ddtwo{the main text.}
As such, we obtain the metamaterial of Fig. \ref{fig_solve7x5}JK, with corresponding modes in Fig. \ref{fig_solve7x5}CFI and Fig. \ref{fig1}EFG in vertex and deformed representation respectively.

\section{Energy in unit cell}
\label{subsec:Energy}
The equations which we use to define the polygons are not exact. For the octagon on a square lattice, they are accurate up to a linear (Eq. \eqref{eq:base1} to Eq. \eqref{eq:base3}) or quadratic order (Eq. \eqref{eq:base1quad} to Eq. \eqref{eq:base3quad}). As the equations are not exact, a solution to these equations will not provide a pure mechanism to exact order.

\begin{figure*}[ht!]
\centering 
\includegraphics[width=1.\textwidth]{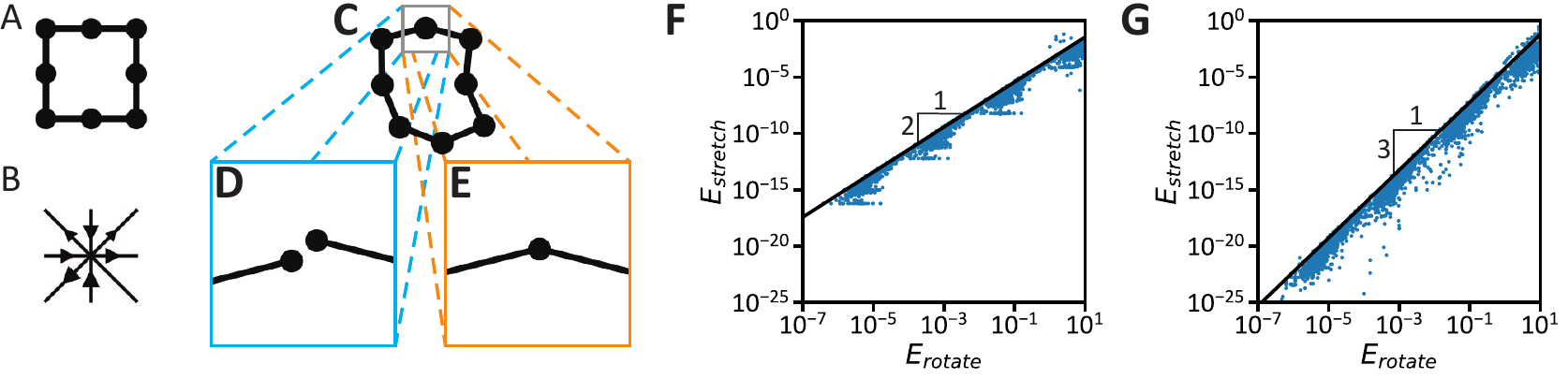}
\caption{\textbf{Unit Cell Energy.} (A) Undeformed Regular Octagon. (BC) Possible deformation mode in vertex (B) and regular (C) representation. (DE) Zoom in top node after rotating all members (D) without and (E) with correction. (FG) Normalized stretch energy to correct deformation $E_{stretch}$  as function of the normalized rotational energy $E_{rotate}$ for (F) linear and (G) quadratically nonlinear analyses.}
\label{fig-energy}
\end{figure*}

This can be demonstrated using Figure \ref{fig-energy}. In Figure \ref{fig-energy}ABC respectively, we see a square octagon in an (A) undeformed configuration, (B) deformed with vertex representation and (C), deformed configuration. Figure \ref{fig-energy}ABC are equivalent to Fig. \ref{fig2}CLF respectively. The deformations of \ref{fig-energy}BC have been obtained by solving Eq. \eqref{eq:base1quad} to Eq. \eqref{eq:base3quad}. Because we do not use an exact approach, we find a mismatch in Fig. \ref{fig-energy}D when we plot the deformation in Fig. \ref{fig-energy}C. We can still connect the nodes by distributing the mismatch evenly across all 8 nodes, as seen in Fig. \ref{fig-energy}E. This would be equivalent to shearing and stretching. We name these deformations $s_{0-7}$. 

We can use this mismatch to quantify the error in our calculations and to increase our understanding of the relation between rotation on one hand and shearing and stretching on the other hand. When a hinge deforms, the strain energy is typically related to the square of the deformation. We can therefore get an an understanding of how strain energy over the unit cells develops by looking at the sum of the squares of the rotations, $E_{rotate}=\sum_{i=0}^{7}a_i$,  and stretches, $E_{stretch}=\sum_{i=0}^{7}s_i$.

To do so, we take the base octagon of Fig. \ref{fig-energy}A and apply five random rotations from $a_{0-7}$. We then calculate the three remaining rotations from $a_{0-7}$ as well as the stretch deformations required $s_{0-7}$, which are all equal. We do this 4000 times for the linear (Eq. \eqref{eq:base1} to Eq. \eqref{eq:base3}) and quadratic solution (Eq. \eqref{eq:base1quad} to Eq. \eqref{eq:base3quad}), spanning a wide range of deformation sizes. We plot the results in Fig. \ref{fig-energy}F for the linear solution and Fig. \ref{fig-energy}G for the quadratic solution. In both cases, we find that the upper boundary of the error can be described using a power law, which has an order of two for the linear solution (Fig. \ref{fig-energy}F) and an order of three for the quadratic solution (Fig. \ref{fig-energy}G). 

This difference between a second and third order order relation, implies that the error decreases much more quickly for small deformations in the quadratic case. At very small deformations ($E_{rotate}=10^{-7}$), the upper level of the error of the quadratic solution is eight orders of magnitude smaller. This shows the improvement of quadratic nonlinearity with respect to a linear solution. 

This assessment clearly shows that the quadratic nonlinearity is more reliable, particularly for small deformations. However, it does not yet specify the degree of stretching or shearing which would be required on a metamaterial level. A variety of methods could be applied to assess this, which is considered out-of-scope for this thesis~\cite{calladine1991first,calladine1992further,luo2006geometrically,muller2009generic}.

\section{Solving polygons}
\label{subsec:solpoly}
In this section, we provide the analytical breakdown of several polygons into three separate equations. While this section shows two specific polygons, a similar approach can be taken for any two-dimensional polygon. The polygons we analyze are shown in Fig. \ref{fig-poly}. In all cases, angles and lengths are denominated by $A$ and $L$, while angular changes with respect to undistorted are denominated by $a$. Displacements of the base nodes in the undeformed case are defined by $d$ in the direction of the orange arrows. The nodes are numbered in black. Angles are subdivided further using the gray and cyan subdivision. $a,A,d$ are denominated by the subscript of the node number, possibly followed by the second subdivision. $L$ are subscripted by the node numbers, which they are linked to. Subscript $b$ indicates undeformed but possibly distorted by $d_i$. It is possible to consider polygons on the same grid with less degrees of freedom by setting $a_i,d_i=0$ for the degrees of freedom which are not to be included.

\begin{figure}[ht!]
\includegraphics[width=\columnwidth]{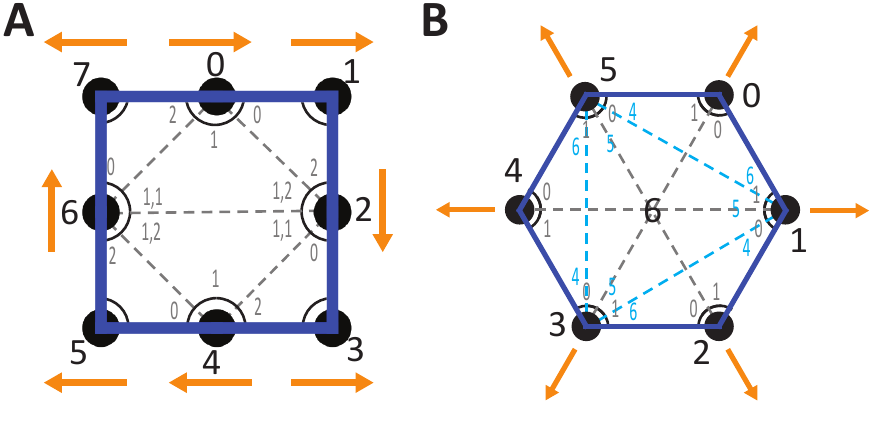}
\caption{\textbf{Split base cell geometries.} (A) Octagon on square lattice. (B) Regular hexagon.}
\label{fig-poly}
\end{figure}

\subsection{Octagonal cell on square lattice}
\label{subsec:App_Deriv_oct}
In this section, we provide the analytical breakdown of an octagonal unit cell on a square lattice, as shown in Fig. \ref{fig-poly}A. This is the cell which has been used throughout the manuscript. This cell is completed defined by 16 variables: $a_i$ and $d_i$, with $i$ between 0 and 7. The cell can be reduced to three equations following the approach given by Eq. \eqref{eq:sqstart} to Eq. \eqref{eq:sq3}. The cell is then defined exclusively by Eq. \eqref{eq:sq1}, Eq. \eqref{eq:sq2} and Eq. \eqref{eq:sq3}, which depend on $a_i$ and $d_i$. Eq. \eqref{eq:base1}, Eq. \eqref{eq:base2} and Eq. \eqref{eq:base3} are the full second order Taylor expansions of Eq. \eqref{eq:sq1}, Eq. \eqref{eq:sq2} and Eq. \eqref{eq:sq3} respectively. It is possible to solve for any three unknowns in $a_i$ and $d_i$, given the other 13 variables by setting Eq. \eqref{eq:sq1}, Eq. \eqref{eq:sq2} and Eq. \eqref{eq:sq3} to 0.

\begin{equation}
{A_{0}}=\pi+{a_{0}}
\label{eq:sqstart}
\end{equation}
\begin{equation}
{A_{1}}=\pi/2+{a_{1}} -\tan^{-1}\left( \frac{d_1}{1+d_2}\right)
\end{equation}
\begin{equation}
{A_{2}}=\pi+{a_{2}}
\end{equation}
\begin{equation}
{A_{3}}=\pi/2+{a_{3}}  -\tan^{-1}\left( \frac{d_3}{1-d_2}\right)
\end{equation}
\begin{equation}
{A_{4}}=\pi+{a_{4}}
\end{equation}
\begin{equation}
{A_{5}}=\pi/2+{a_{5}}  -\tan^{-1}\left( \frac{d_5}{1+d_6}\right)
\end{equation}
\begin{equation}
{A_{6}}=\pi+{a_{6}}
\end{equation}
\begin{equation}
{A_{7}}=\pi/2+{a_{7}}  -\tan^{-1}\left( \frac{d_7}{1-d_6}\right)
\end{equation}

\begin{equation}
{L_{7-0}} = 1+{d_{0}}+{d_{7}}
\end{equation}
\begin{equation}
{L_{0-1}} = 1-{d_{0}}+{d_{1}}
\end{equation}
\begin{equation}
{L_{1-2}} = \sqrt{(  (1+{d_{2}})^2 + {d_{1}}^2)}
\end{equation}
\begin{equation}
{L_{2-3}} = \sqrt{(  (1-{d_{2}})^2 + {d_{3}}^2)}
\end{equation}
\begin{equation}
{L_{3-4}} = 1+{d_{4}}+{d_{3}}
\end{equation}
\begin{equation}
{L_{4-5}} = 1-{d_{4}}+{d_{5}}
\end{equation}
\begin{equation}
{L_{5-6}} = \sqrt{(  (1+{d_{6}})^2 + {d_{5}}^2)}
\end{equation}
\begin{equation}
{L_{6-7}} = \sqrt{(  (1-{d_{6}})^2 + {d_{7}}^2)}
\end{equation}
\begin{equation}
\end{equation}

\small
\begin{equation}
{L_{0-2}} = \sqrt{({L_{0-1}}^2+{L_{1-2}}^2-2 \cdot {L_{0-1}} \cdot {L_{1-2}} \cdot  \cos ({A_{1}}))}
\end{equation}
\begin{equation}
{L_{2-4}} = \sqrt{({L_{2-3}}^2+{L_{3-4}}^2-2 \cdot {L_{2-3}} \cdot {L_{3-4}} \cdot  \cos ({A_{3}}))}
\end{equation}
\begin{equation}
{L_{4-6}} = \sqrt{({L_{4-5}}^2+{L_{5-6}}^2-2 \cdot {L_{4-5}} \cdot {L_{5-6}} \cdot  \cos ({A_{5}}))}
\end{equation}
\begin{equation}
{L_{6,0}} = \sqrt{({L_{6-7}}^2+{L_{7-0}}^2-2 \cdot {L_{6-7}} \cdot {L_{7-0}} \cdot  \cos ({A_{7}}))}
\end{equation}
\normalsize

\begin{equation}
{A_{0,0}} = \sin^{-1} ( {L_{1-2}}/{L_{0-2}}  \cdot  \sin({A_{1}}) )
\end{equation}
\begin{equation}
{A_{2,2}} = \pi-{A_{1}}-{A_{0,0}} 
\end{equation}
\begin{equation}
{A_{2,0}} = \sin^{-1} ( {L_{3-4}}/{L_{2-4}}  \cdot  \sin({A_{3}}) )
\end{equation}
\begin{equation}
{A_{4,2}} = \pi-{A_{3}}-{A_{2,0}} 
\end{equation}
\begin{equation}
{A_{4,0}} = \sin^{-1} ( {L_{5-6}}/{L_{4-6}}  \cdot  \sin({A_{5}}) )
\end{equation}
\begin{equation}
{A_{6,2}} = \pi-{A_{5}}-{A_{4,0}} 
\end{equation}
\begin{equation}
{A_{6,0}} = \sin^{-1} ( {L_{7-0}}/{L_{6,0}}  \cdot  \sin({A_{7}}) )
\end{equation}
\begin{equation}
{A_{0,2}} = \pi-{A_{7}}-{A_{6,0}} 
\end{equation}
\begin{equation}
{A_{0,1}} = {A_{0}}-{A_{0,0}}-{A_{0,2}}
\end{equation}

\small
\begin{equation}
{L_{2-6}} = \sqrt{({L_{0-2}}^2+{L_{6,0}}^2-2 \cdot {L_{0-2}} \cdot {L_{6,0}} \cdot  \cos ({A_{0,1}}))}
\end{equation}
\begin{equation}
{A_{4,1}} = \cos^{-1} (({L_{2-4}}^2+{L_{4-6}}^2-{L_{2-6}}^2)/(2 \cdot {L_{2-4}} \cdot {L_{4-6}}))
\end{equation}
\normalsize

\begin{equation}
{A_{4}} = {A_{4,0}}+{A_{4,1}}+{A_{4,2}} 
\end{equation}
\begin{equation}
{A_{6,1,1}} = \sin^{-1} (\sin({A_{0,1}}) \cdot {L_{0-2}}/{L_{2-6}})
\end{equation}
\begin{equation}
{A_{2,1,2}} = \sin^{-1} (\sin({A_{0,1}}) \cdot {L_{6,0}}/{L_{2-6}})
\end{equation}
\begin{equation}
{A_{6,1,2}} = \sin^{-1} (\sin({A_{4,1}}) \cdot {L_{2-4}}/{L_{2-6}})
\end{equation}
\begin{equation}
{A_{2,1,1}} = \sin^{-1} (\sin({A_{4,1}}) \cdot {L_{4-6}}/{L_{2-6}})
\end{equation}
\begin{equation}
{A_{2}} = {A_{2,0}}+{A_{2,1,1}}+{A_{2,1,2}}+{A_{2,2}}
\end{equation}
\begin{equation}
{A_{6}} = {A_{6,0}}+{A_{6,1,1}}+{A_{6,1,2}}+{A_{6,2}}
\end{equation}

\begin{equation}
{a_{2}}_{sol}={A_{2}}-\pi 
\end{equation}
\begin{equation}
{a_{4}}_{sol}={A_{4}}-\pi 
\end{equation}
\begin{equation}
{a_{6}}_{sol}={A_{6}}-\pi 
\end{equation}

\begin{equation}
eq_{1} = {a_{2}}-{a_{2}}_{sol}
\label{eq:sq1}
\end{equation}
\begin{equation}
eq_{2} = {a_{4}}-{a_{4}}_{sol}
\label{eq:sq2}
\end{equation}
\begin{equation}
eq_{3} = {a_{6}}-{a_{6}}_{sol}
\label{eq:sq3}
\end{equation}

Eq. \eqref{eq:sq1} to \eqref{eq:sq3} can be exactly solved and reduced to the following quadratically nonlinear equations:

\scriptsize
\begin{equation}
\begin{aligned}
0=& - a_{0} - \frac{3 a_{1}}{2} - a_{2}- \frac{a_{3}}{2}  + \frac{a_{5}}{2}	- \frac{a_{7}}{2} + d_{1}  + d_{3} 	\\
& \blue{\frac{a_{0} a_{3}}{2}} - \blue{\frac{a_{0} a_{7}}{2}} - \blue{\frac{a_{0} d_{0}}{2}} - \blue{\frac{a_{0} d_{2}}{2}} + \blue{\frac{a_{0} d_{4}}{2}} + \\
&\blue{\frac{a_{0} d_{6}}{2}}  + \blue{\frac{a_{1}^{2}}{8}} + \blue{\frac{a_{1} a_{3}}{2}} - \blue{\frac{a_{1} a_{7}}{4}} - \blue{\frac{a_{1} d_{1}}{2}} - \\
&\blue{\frac{3 a_{1} d_{2}}{4}} + \blue{\frac{a_{1} d_{4}}{2}} + \blue{\frac{a_{1} d_{6}}{4}} - \blue{\frac{a_{3}^{2}}{8}} - \\
&\blue{\frac{a_{3} a_{5}}{4}} + \blue{\frac{a_{3} a_{7}}{2}} - \blue{\frac{a_{3} d_{2}}{4}} - \blue{\frac{a_{3} d_{3}}{2}} - \blue{\frac{a_{3} d_{4}}{2}} - \\
&\blue{\frac{a_{3} d_{6}}{4}} - \blue{\frac{a_{5}^{2}}{8}} + \blue{\frac{a_{5} d_{2}}{4}} - \blue{\frac{a_{5} d_{4}}{2}} + \\
&\blue{\frac{a_{5} d_{5}}{2}} + \blue{\frac{a_{5} d_{6}}{4}}  - \blue{\frac{3 a_{7}^{2}}{8}} - \blue{\frac{a_{7} d_{2}}{4}} + \\
&\blue{\frac{a_{7} d_{4}}{2}} + \blue{\frac{3 a_{7} d_{6}}{4}} + \blue{\frac{a_{7} d_{7}}{2}} - \blue{d_{1} d_{2}} +\blue{d_{2} d_{3}},
\end{aligned}
\label{eq:base1quad}
\end{equation}

\begin{equation}
\begin{aligned}
0=& + {a_{0}} + {a_{1}}- {a_{3}} - {a_{4}} -  {a_{5}}  + a_{7}\\
&\blue{\frac{a_{0} a_{1}}{2}} - \blue{\frac{a_{0} a_{3}}{2}} - \blue{\frac{a_{0} a_{5}}{2}} + \blue{\frac{a_{0} a_{7}}{2}} + \blue{a_{0} d_{2}} - \blue{a_{0} d_{6}} + \\
& \blue{\frac{a_{1}^{2}}{4}} - \blue{\frac{a_{1} a_{3}}{2}} - \blue{\frac{a_{1} a_{5}}{2}} + \blue{\frac{a_{1} a_{7}}{2}} + \blue{\frac{3 a_{1} d_{2}}{2}} - \\
&\blue{\frac{a_{1} d_{6}}{2}}  + \blue{\frac{a_{3}^{2}}{4}} + \blue{\frac{a_{3} a_{5}}{2}} - \blue{\frac{a_{3} a_{7}}{2}} + \blue{\frac{a_{3} d_{2}}{2}} + \\
&\blue{\frac{a_{3} d_{6}}{2}}  + \blue{\frac{a_{5}^{2}}{4}} - \blue{\frac{a_{5} a_{7}}{2}} - \blue{\frac{a_{5} d_{2}}{2}} - \blue{\frac{a_{5} d_{6}}{2}} + \\
& \blue{\frac{a_{7}^{2}}{4}} + \blue{\frac{a_{7} d_{2}}{2}} - \blue{\frac{3 a_{7} d_{6}}{2}},
\end{aligned}
\label{eq:base2quad}
\end{equation}

\begin{equation}
\begin{aligned}
0 = & - {a_{0}} - {\frac{a_{1}}{2}}+ {\frac{a_{3}}{2}}  - {\frac{a_{5}}{2}} - {a_{6}} - {\frac{3 a_{7}}{2}} + {d_{5}} + {d_{7}} \\
& - \blue{\frac{a_{0} a_{1}}{2}} + \blue{\frac{a_{0} a_{5}}{2}} + \blue{\frac{a_{0} d_{0}}{2}} - \blue{\frac{a_{0} d_{2}}{2}} - \blue{\frac{a_{0} d_{4}}{2}} + \\
&\blue{\frac{a_{0} d_{6}}{2}}  - \blue{\frac{3 a_{1}^{2}}{8}} + \blue{\frac{a_{1} a_{5}}{2}} - \blue{\frac{a_{1} a_{7}}{4}} + \blue{\frac{a_{1} d_{1}}{2}} - \\
&\blue{\frac{3 a_{1} d_{2}}{4}} - \blue{\frac{a_{1} d_{4}}{2}} + \blue{\frac{a_{1} d_{6}}{4}}  - \blue{\frac{a_{3}^{2}}{8}} - \blue{\frac{a_{3} a_{5}}{4}} - \\
&\blue{\frac{a_{3} d_{2}}{4}} + \blue{\frac{a_{3} d_{3}}{2}} + \blue{\frac{a_{3} d_{4}}{2}} - \blue{\frac{a_{3} d_{6}}{4}} - \blue{\frac{a_{5}^{2}}{8}} + \\
&\blue{\frac{a_{5} a_{7}}{2}} + \blue{\frac{a_{5} d_{2}}{4}} + \blue{\frac{a_{5} d_{4}}{2}} - \blue{\frac{a_{5} d_{5}}{2}} + \blue{\frac{a_{5} d_{6}}{4}} + \\
& \blue{\frac{a_{7}^{2}}{8}} - \blue{\frac{a_{7} d_{2}}{4}} - \blue{\frac{a_{7} d_{4}}{2}} + \blue{\frac{3 a_{7} d_{6}}{4}} - \blue{\frac{a_{7} d_{7}}{2}} - \\
& \blue{d_{5} d_{6}} + \blue{d_{6} d_{7}},
\end{aligned}
\label{eq:base3quad}
\end{equation}
\normalsize

where linear terms are highlighted in black and quadratic terms in blue. The linear solution without quadratic terms is give in Eq. \eqref{eq:base1} to Eq. \eqref{eq:base3}.

\subsection{Hexagonal cell}
\label{subsec:hexcell}
In this section, we provide the analytical breakdown of a regular hexagonal unit cell, as shown in Fig. \ref{fig-poly}B. This cell is completed defined by 12 variables: $a_i$ and $d_i$, with $i$ between 0 and 5. The cell can be reduced to three equations following the approach given by Eq. \eqref{eq:hexstart} to Eq. \eqref{eq:hex3}. The cell is then defined exclusively by Eq. \eqref{eq:hex1}, Eq. \eqref{eq:hex2} and Eq. \eqref{eq:hex3}, which depend on $a_i$ and $d_i$. It is possible to solve for any three unknowns in $a_i$ and $d_i$, given the other 9 variables by setting Eq. \eqref{eq:hex1}, Eq. \eqref{eq:hex2} and Eq. \eqref{eq:hex3} to 0.

\begin{equation}
{A_b} = \pi/3
\label{eq:hexstart}
\end{equation}
\begin{equation}
{L_{0-6,b}} = 1+d0
\end{equation}
\begin{equation}
{L_{1-6,b}} = 1+d1
\end{equation}
\begin{equation}
{L_{2-6,b}} = 1+d2
\end{equation}
\begin{equation}
{L_{3-6,b}} = 1+d3
\end{equation}
\begin{equation}
{L_{4-6,b}} = 1+d4
\end{equation}
\begin{equation}
{L_{5-6,b}} = 1+d5
\end{equation}

\scriptsize
\begin{equation}
{L_{0-1}} = \sqrt{ {L_{0-6,b}}^2+{L_{1-6,b}}^2 -2 \cdot {L_{0-6,b}} \cdot {L_{1-6,b}} \cdot cos({A_b})) }
\end{equation}
\begin{equation}
{L_{1-2}} = \sqrt{ {L_{1-6,b}}^2+{L_{2-6,b}}^2 -2 \cdot {L_{1-6,b}} \cdot {L_{2-6,b}} \cdot cos({A_b})) }
\end{equation}
\begin{equation}
{L_{2-3}} = \sqrt{ {L_{2-6,b}}^2+{L_{3-6,b}}^2 -2 \cdot {L_{2-6,b}} \cdot {L_{3-6,b}} \cdot cos({A_b})) }
\end{equation}
\begin{equation}
{L_{3-4}} = \sqrt{ {L_{3-6,b}}^2+{L_{4-6,b}}^2 -2 \cdot {L_{3-6,b}} \cdot {L_{4-6,b}} \cdot cos({A_b})) }
\end{equation}
\begin{equation}
{L_{4-5}} = \sqrt{ {L_{4-6,b}}^2+{L_{5-6,b}}^2 -2 \cdot {L_{4-6,b}} \cdot {L_{5-6,b}} \cdot cos({A_b})) }
\end{equation}
\begin{equation}
{L_{5-0}} = \sqrt{ {L_{5-6,b}}^2+{L_{0-6,b}}^2 -2 \cdot {L_{5-6,b}} \cdot {L_{0-6,b}} \cdot cos({A_b})) }
\end{equation}
\normalsize

\begin{equation}
{A_{0,0,b}} = {A_b}/{L_{0-1}}  \cdot  {L_{1-6,b}}
\end{equation}
\begin{equation}
{A_{0,1,b}} = {A_b}/{L_{5-0}}  \cdot  {L_{5-6,b}}
\end{equation}
\begin{equation}
{A_{0,b}} = {A_{0,0,b}}+{A_{0,1,b}}
\end{equation}

\begin{equation}
{A_{1,0,b}} = {A_b}/{L_{1-2}}  \cdot  {L_{2-6,b}}
\end{equation}
\begin{equation}
{A_{1,1,b}} = {A_b}/{L_{0-1}}  \cdot  {L_{0-6,b}}
\end{equation}
\begin{equation}
{A_{1,b}} = {A_{1,0,b}}+{A_{1,1,b}}
\end{equation}

\begin{equation}
{A_{2,0,b}} = {A_b}/{L_{2-3}}  \cdot  {L_{3-6,b}}
\end{equation}
\begin{equation}
{A_{2,1,b}} = {A_b}/{L_{1-2}}  \cdot  {L_{1-6,b}}
\end{equation}
\begin{equation}
{A_{2,b}} = {A_{2,0,b}}+{A_{2,1,b}}
\end{equation}
\begin{equation}
{A_{3,0,b}} = {A_b}/{L_{3-4}}  \cdot  {L_{4-6,b}}
\end{equation}
\begin{equation}
{A_{3,1,b}} = {A_b}/{L_{2-3}}  \cdot  {L_{2-6,b}}
\end{equation}
\begin{equation}
{A_{3,b}} = {A_{3,0,b}}+{A_{3,1,b}}
\end{equation}
\begin{equation}
{A_{4,0,b}} = {A_b}/{L_{4-5}}  \cdot  {L_{5-6,b}}
\end{equation}
\begin{equation}
{A_{4,1,b}} = {A_b}/{L_{3-4}}  \cdot  {L_{3-6,b}}
\end{equation}
\begin{equation}
{A_{4,b}} = {A_{4,0,b}}+{A_{4,1,b}}
\end{equation}
\begin{equation}
{A_{5,0,b}} = {A_b}/{L_{5-0}}  \cdot  {L_{0-6,b}}
\end{equation}
\begin{equation}
{A_{5,1,b}} = {A_b}/{L_{4-5}}  \cdot  {L_{4-6,b}}
\end{equation}
\begin{equation}
{A_{5,b}} = {A_{5,0,b}}+{A_{5,1,b}}
\end{equation}

\begin{equation}
{A_0}={A_{0,b}}+a0
\end{equation}
\begin{equation}
{A_2}={A_{2,b}}+{a_{2}}
\end{equation}
\begin{equation}
{A_4}={A_{4,b}}+{a_{4}}
\end{equation}

\small
\begin{equation}
{L_{1-5}} = \sqrt{ {L_{0-1}}^2 + {L_{5-0}}^2 - 2 \cdot {L_{0-1}} \cdot {L_{5-0}} \cdot cos({A_0}))}
\end{equation}
\begin{equation}
{L_{1-3}} = \sqrt{ {L_{1-2}}^2 + {L_{2-3}}^2 - 2 \cdot {L_{1-2}} \cdot {L_{2-3}} \cdot cos({A_2}))}
\end{equation}
\begin{equation}
{L_{3-5}} = \sqrt{ {L_{3-4}}^2 + {L_{4-5}}^2 - 2 \cdot {L_{3-4}} \cdot {L_{4-5}} \cdot cos({A_4}))}
\end{equation}
\normalsize

\begin{equation}
{A_{1,4}} = {A_2}/{L_{1-3}}  \cdot  {L_{2-3}} 
\end{equation}
\begin{equation}
{A_{1,5}} = \cos^{-1} \left( \frac{{L_{1-5}}^2+{L_{1-3}}^2-{L_{3-5}}^2}{2 \cdot {L_{1-5}} \cdot {L_{1-3}}} \right)
\end{equation}
\begin{equation}
{A_{1,6}} = {A_0}/{L_{1-5}}  \cdot  {L_{5-0}}
\end{equation}
\begin{equation}
{A_{1}} = {A_{1,4}}+{A_{1,5}}+{A_{1,6}}
\end{equation}
\begin{equation}
{a_{1,sol}}={A_{1}}-{A_{1,b}}
\end{equation}

\begin{equation}
{A_{3,4}} = {A_4}/{L_{3-5}}  \cdot  {L_{4-5}} 
\end{equation}
\begin{equation}
{A_{3,5}} = \cos^{-1} \left(\frac{{L_{1-3}}^2+{L_{3-5}}^2-{L_{1-5}}^2}{2 \cdot {L_{3-5}} \cdot {L_{1-3}}}\right)
\end{equation}
\begin{equation}
{A_{3,6}} = {A_2}/{L_{1-3}}  \cdot  {L_{1-2}}
\end{equation}
\begin{equation}
{A_{3}} = {A_{3,4}}+{A_{3,5}}+{A_{3,6}}
\end{equation}
\begin{equation}
{a_{3,sol}}={A_{3}}-{A_{3,b}}
\end{equation}

\begin{equation}
{A_{5,4}} = {A_0}/{L_{1-5}}  \cdot  {L_{0-1}} 
\end{equation}
\begin{equation}
{A_{5,5}} = \cos^{-1} \left( \frac{{L_{1-5}}^2+{L_{3-5}}^2-{L_{1-3}}^2}{2 \cdot {L_{1-5}} \cdot {L_{3-5}}} \right)
\end{equation}
\begin{equation}
{A_{5,6}} = {A_4}/{L_{3-5}}  \cdot  {L_{3-4}}
\end{equation}
\begin{equation}
{A_{5}} = {A_{5,4}}+{A_{5,5}}+{A_{5,6}}
\end{equation}
\begin{equation}
{a_{5,sol}}={A_{5}}-{A_{5,b}}
\end{equation}

\begin{equation}
eq_{1,h}= {a_{1}}-{a_{1,sol}} 
\label{eq:hex1}
\end{equation}
\begin{equation}
eq_{2,h}= {a_{3}}-{a_{3,sol}} 
\label{eq:hex2}
\end{equation}
\begin{equation}
eq_{3,h}= {a_{5}}-{a_{5,sol}} 
\label{eq:hex3}
\end{equation}

Eq. \eqref{eq:hex1} to \eqref{eq:hex3} can be exactly solved and reduced to the following linear equations:

\begin{equation}
\begin{aligned}
0= & - \frac{\pi a_{0}}{9} - \frac{a_{0}}{6} + \frac{\sqrt{3} a_{0}}{3} - a_{1} - \frac{\pi a_{2}}{9} - \frac{a_{2}}{6} + \\
&\frac{\sqrt{3} a_{2}}{3} + \frac{a_{4}}{3} - \frac{\sqrt{3} \pi d_{0}}{9} - \frac{\pi d_{0}}{9} - \frac{\sqrt{3} d_{0}}{6} + \\
&\frac{\pi^{2} d_{0}}{27} - \frac{\pi^{2} d_{1}}{27} - \frac{\sqrt{3} d_{1}}{6} + \frac{5 \pi d_{1}}{18} - \frac{\sqrt{3} \pi d_{2}}{9} - \\
&\frac{\pi d_{2}}{9} - \frac{\sqrt{3} d_{2}}{6} + \frac{\pi^{2} d_{2}}{27} - \frac{\pi^{2} d_{3}}{54} + \frac{\pi d_{3}}{36} + \\
&\frac{\sqrt{3} d_{3}}{12} + \frac{\sqrt{3} \pi d_{3}}{9} - \frac{\pi d_{4}}{9} + \frac{\sqrt{3} d_{4}}{3} - \frac{\pi^{2} d_{5}}{54} + \\
&\frac{\pi d_{5}}{36} + \frac{\sqrt{3} d_{5}}{12} + \frac{\sqrt{3} \pi d_{5}}{9} - \frac{\pi}{3} + \frac{4 \sqrt{3} \pi}{9},
\end{aligned}
\label{eq:hex1lin}
\end{equation}

\begin{equation}
\begin{aligned}
0 =& \frac{a_{0}}{3} - \frac{\pi a_{2}}{9} - \frac{a_{2}}{6} + \frac{\sqrt{3} a_{2}}{3} - a_{3} - \frac{\pi a_{4}}{9} - \\
&\frac{a_{4}}{6} + \frac{\sqrt{3} a_{4}}{3} - \frac{\pi d_{0}}{9} + \frac{\sqrt{3} d_{0}}{3} - \frac{\pi^{2} d_{1}}{54} + \\
&\frac{\pi d_{1}}{36} + \frac{\sqrt{3} d_{1}}{12} + \frac{\sqrt{3} \pi d_{1}}{9} - \frac{\sqrt{3} \pi d_{2}}{9} - \frac{\pi d_{2}}{9} - \\
&\frac{\sqrt{3} d_{2}}{6} + \frac{\pi^{2} d_{2}}{27} - \frac{\pi^{2} d_{3}}{27} - \frac{\sqrt{3} d_{3}}{6} + \frac{5 \pi d_{3}}{18} - \\
&\frac{\sqrt{3} \pi d_{4}}{9} - \frac{\pi d_{4}}{9} - \frac{\sqrt{3} d_{4}}{6} + \frac{\pi^{2} d_{4}}{27} - \frac{\pi^{2} d_{5}}{54} + \\
&\frac{\pi d_{5}}{36} + \frac{\sqrt{3} d_{5}}{12} + \frac{\sqrt{3} \pi d_{5}}{9} - \frac{\pi}{3} + \frac{4 \sqrt{3} \pi}{9},
\end{aligned}
\label{eq:hex2lin}
\end{equation}

\begin{equation}
\begin{aligned}
0=& - \frac{\pi a_{0}}{9} - \frac{a_{0}}{6} + \frac{\sqrt{3} a_{0}}{3} + \frac{a_{2}}{3} - \frac{\pi a_{4}}{9} - \frac{a_{4}}{6} + \\
&\frac{\sqrt{3} a_{4}}{3} - a_{5} - \frac{\sqrt{3} \pi d_{0}}{9} - \frac{\pi d_{0}}{9} - \frac{\sqrt{3} d_{0}}{6} + \frac{\pi^{2} d_{0}}{27} - \\
&\frac{\pi^{2} d_{1}}{54} + \frac{\pi d_{1}}{36} + \frac{\sqrt{3} d_{1}}{12} + \frac{\sqrt{3} \pi d_{1}}{9} - \frac{\pi d_{2}}{9} + \\
&\frac{\sqrt{3} d_{2}}{3} - \frac{\pi^{2} d_{3}}{54} + \frac{\pi d_{3}}{36} + \frac{\sqrt{3} d_{3}}{12} + \frac{\sqrt{3} \pi d_{3}}{9} - \\
&\frac{\sqrt{3} \pi d_{4}}{9} - \frac{\pi d_{4}}{9} - \frac{\sqrt{3} d_{4}}{6} + \frac{\pi^{2} d_{4}}{27} - \frac{\pi^{2} d_{5}}{27} - \\
&\frac{\sqrt{3} d_{5}}{6} + \frac{5 \pi d_{5}}{18} - \frac{\pi}{3} + \frac{4 \sqrt{3} \pi}{9}.
\end{aligned}
\label{eq:hex3lin}
\end{equation}

This solution is significantly longer than Eq. \eqref{eq:base1} to Eq. \eqref{eq:base3} as there are no right or straight angles so less terms cancel out.

\end{appendix} 

\end{document}